# Dirac cones reshaped by interaction effects in suspended graphene

D. C. Elias[1], R. V. Gorbachev[1], A. S. Mayorov[1], S. V. Morozov[2], A. A. Zhukov[3],
P. Blake[3], L. A. Ponomarenko[1], I. V. Grigorieva[1], K. S. Novoselov[1], F. Guinea[4], A. K. Geim[1,3]

[1]School of Physics & Astronomy, University of Manchester, Manchester M13 9PL, UK
[2]Institute for Microelectronics Technology, 142432 Chernogolovka, Russia
[3]Manchester Centre for Mesoscience & Nanotechnology, University of Manchester, Manchester M13 9PL, UK
[4]Instituto de Ciencia de Materiales de Madrid (CSIC), Sor Juana Inés de la Cruz 3, Madrid 28049, Spain

We report measurements of the cyclotron mass in graphene for carrier concentrations $n$ varying over three orders of magnitude. In contrast to the single-particle picture, the real spectrum of graphene is profoundly nonlinear so that the Fermi velocity describing the spectral slope reaches $\approx 3 \times 10^6$ m/s at $n < 10^{10}$ cm$^{-2}$, three times the value commonly used for graphene. The observed changes are attributed to electron-electron interaction that renormalizes the Dirac spectrum because of weak screening. Our experiments also put an upper limit of ~0.1 meV on the possible gap in graphene.

In graphene, electron-electron interactions are expected to play a significant role as the screening length diverges at the charge neutrality point (NP) and the conventional Landau theory that allows one to map a strongly-interacting electronic liquid into a gas of non-interacting fermions is no longer applicable [1,2]. This should result in considerable changes in graphene's linear spectrum, and even more dramatic scenarios including the opening of an energy gap have also been proposed [3-5]. Experimental evidence for such spectral changes is scarce, so that its strongest piece is probably a 20% difference between the Fermi velocities $v_F$ found in graphene and carbon nanotubes [6]. In this Letter, we report measurements of the cyclotron mass $m_c$ for a wide range of $n$ and show that the widely-used linear approximation for graphene's spectrum is no longer valid at low $n$. Our experiments yield a pronounced enhancement in $v_F$, which is logarithmic in $n$ and can be described by the renormalization group theory (RGT) [1].

In the first approximation, charge carriers in graphene behave like massless relativistic particles with a conical energy spectrum $E = v_F \hbar k$ where the Fermi velocity $v_F$ plays the role of the effective speed of light and $k$ is the wavevector. Because graphene's spectrum is filled with electronic states up to the Fermi energy, their Coulomb interaction has to be taken into account. To do this, the standard approach of Landau's Fermi-liquid theory proven successful for normal metals fails in graphene, especially at $E$ close to the NP where the density of states vanishes. This leads to theoretical divergences that have the same origin as those in quantum electrodynamics and other interacting field theories. In the latter case, the interactions are normally accounted for by using RGT, that is, by defining effective models with a reduced number of degrees of freedom and treating the effect of high-energy excitations perturbatively. This approach was also applied to graphene by using as a small parameter either the effective coupling constant $\alpha = e^2/\hbar v_F$ [7,8] or the inverse of the number of fermion species in graphene $N_f = 4$ [9,10]. The resulting many-body spectrum is shown in Fig. 1. One can see that electron-electron (e-e) interactions reduce the density of states at low $E$ and lead to an increase in $v_F$ that slowly (logarithmically) diverges at zero $E$.

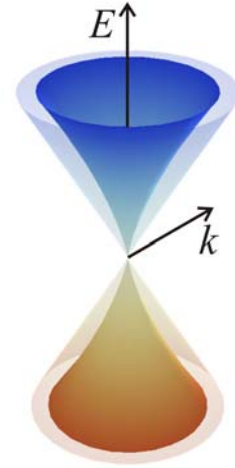

FIG. 1. Sketch of graphene's electronic spectrum with and without taking into account e-e interactions. The outer cone is the single-particle spectrum $E = v_F \hbar k$, and the inner cone illustrates the many-body spectrum predicted by the RGT and observed in the experiments described in this report.

As for experiment, graphene placed on top of an oxidized Si wafer and having typical $n \approx 10^{12}$ cm$^{-2}$ exhibits $v_F = v^*_F \approx 1.05 \pm 0.1 \times 10^6$ m/s. This value was measured by using a variety of techniques including the early transport experiments, in which Shubnikov-de Haas oscillations (SdHO) were analyzed to extract $v_F$ [11,12]. It has been noted that $v^*_F$ is larger than $v^0_F \approx 0.85 \pm 0.05 \times 10^6$ m/s, the value accepted for metallic carbon nanotubes (see, e.g., Ref. 6). In agreement with this notion, the energy gaps measured in semiconducting nanotubes show a nonlinear dependence on their inverse radii, which is consistent with the larger $v_F$ in flat graphene [6]. The differences between $v_F$ in graphene and its rolled-up version can be attributed to e-e interactions [13]. The most direct piece of evidence obtained so far comes from infrared measurements of the Pauli blocking in graphene, which showed a small but sharp (15%) decrease in $v_F$ with increasing $n$ from $\approx 0.5$ to $2 \times 10^{12}$ cm$^{-2}$ [14]. However, the error bars in this experiment were also large ($\approx 10\%$).

To address the discussed problem of "missing" e-e interactions, we have studied SdHO in suspended graphene devices (inset in Fig. 2a). They were fabricated by using the procedures described previously [15-17]. After current annealing, our devices exhibited record mobilities $\mu \sim 1{,}000{,}000$ cm$^2$/Vs, and charge homogeneity $\delta n$ was better than $10^9$ cm$^{-2}$ so that we observed the onset of SdHO in magnetic fields $B \approx 0.01$T and the first quantum Hall plateau became clearly visible in $B$ below 0.1T (see Supplementary Information [18]). To extract the information about graphene's electronic spectrum, we employed the following routine. SdHO were measured at various $B$ and $n$ as a function of temperature ($T$). Their amplitude was then analyzed by using the standard Lifshitz-Kosevich (LK) formula $T/\sinh(2\pi^2 T m_c/\hbar eB)$, which holds for the Dirac spectrum [19] and allows one to find the effective cyclotron mass $m_c$ at a given $n$. This approach was previously employed for graphene on SiO$_2$, and it was shown that, within experimental accuracy and for a range of $n \sim 10^{12}$ cm$^{-2}$, $m_c$ was well described by dependence $m_c = \hbar(\pi n)^{1/2}/v^*_F$, which corresponds to the linear spectrum [11,12]. With respect to the earlier experiments, our suspended devices offer critical advantages. First, in the absence of a substrate, interaction-induced spectral changes are expected to be maximal because no dielectric screening is present. Second, the high quality of suspended graphene has allowed us to probe its spectrum over a wide range of $n$ well below $10^{12}$ cm$^{-2}$, which is essential as the spectral changes are expected to be logarithmic in $n$. Third, due to low $\delta n$, we can approach the Dirac point within a few meV. This low-$E$ regime, in which a major renormalization of the Dirac spectrum is expected, has previously been inaccessible.

Figure 2a plots examples of $T$ dependence of the SdHO amplitude at low $n$ (for further examples of SdHO and their $T$ dependence, see [18]). The curves are well described by the LK formula but the inferred $m_c$ are twice lower than expected if we assume that $v_F$ retains its conventional value $v^*_F$. To emphasize this profound discrepancy with the earlier experiments, the dashed curves in Fig. 2a plot the $T$ dependence expected under the assumption $v_F = v^*_F$. The SdHO would then have to decay twice faster with increasing $T$, which would result in a qualitatively different behavior of SdHO. From the measured $m_c$ we find $v_F \approx 1.9$ and $2.2 \times 10^6$ m/s for the higher and lower $|n|$ in Fig. 2a, respectively. We have carried out measurements of $m_c$ such as in Fig. 2a for many different $n$, and the extracted values are presented in Fig. 2b for one of the devices. For the linear spectrum, $m_c$ is expected to increase linearly with $k_F = (\pi n)^{1/2}$. In contrast, the experiment clearly shows a super-linear behavior. Trying to fit the experimental curves in Fig. 2b with the linear dependence $m_c(k_F)$, we find $v_F \geq 2.5 \times 10^6$ m/s at $n < 10^{10}$ cm$^{-2}$ and $\leq 1.5 \times 10^6$ m/s for $n > 2 \times 10^{11}$ cm$^{-2}$ as indicated by the dashed lines. The observed super-linear dependence of $m_c$ can be translated into $v_F$ varying with $n$. Fig. 2c replots the data in Fig. 2b in terms of $v_F = \hbar(\pi n)^{1/2}/m_c$ which shows a diverging-like behavior of $v_F$ near the NP. This sharp increase in $v_F$ (by nearly a factor of 3 with respect to $v^*_F$) contradicts to the linear model of graphene's spectrum but is consistent with the spectrum reshaped by e-e interactions (Fig. 1).

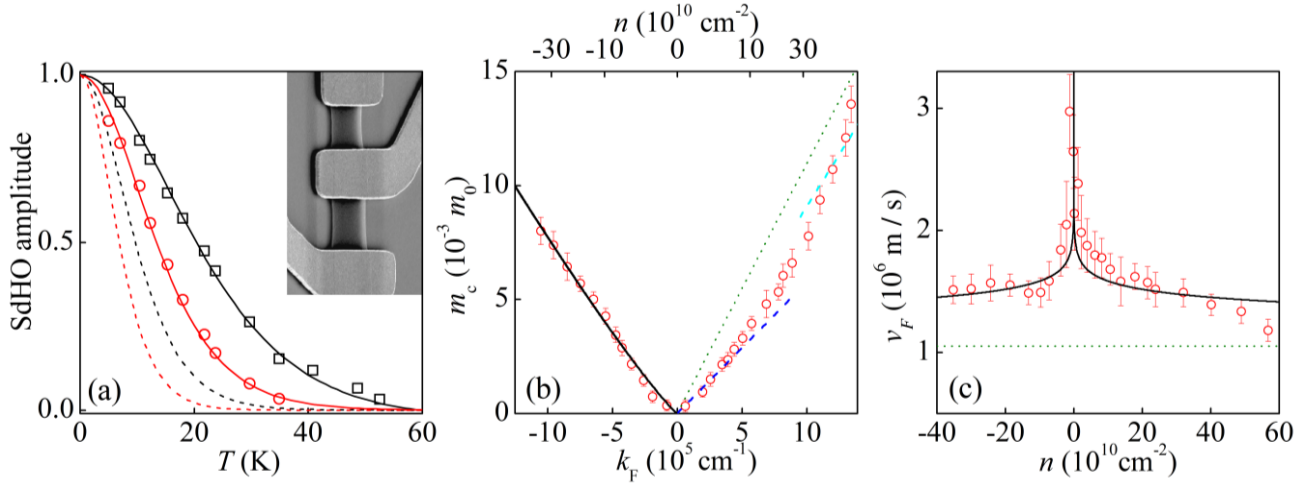

FIG. 2. Probing graphene's electronic spectrum through SdHO. (a) Symbols show examples of the $T$ dependence of SdHO for $n \approx +1.4$ and $-7.0 \times 10^{10}$ cm$^{-2}$ where the sign $\pm$ corresponds to electrons and holes, respectively. The dependence is well described by the LK formula (solid curves). The dashed curves are the behavior expected for $v_F = v^*_F$ (in the matching colors). The inset shows a scanning electron micrograph of one of our devices. The vertical graphene wire is $\approx 2$ μm wide and suspended above an oxidized Si wafer being attached to Au/Cr contacts. Approximately a half of 300 nm thick SiO$_2$ was etched away underneath the graphene structure. (b) $m_c$ as a function of $k_F$ for the same device. The exponential dependence of SdHO's amplitude on $m_c$ allows high accuracy in determining the cyclotron mass, as shown by the error bars. The dashed curves are the best linear fits $m_c \propto n^{1/2}$ at high and low $n$. The dotted line is the behaviour of $m_c$ expected for the standard value of $v_F = v^*_F$. Graphene's spectrum renormalized due to e-e interactions is expected to result in the dependence shown by the solid curve. (c) $m_c$ re-plotted in terms of varying $v_F$. The color scheme is to match the corresponding data in (b).

The data for $m_c$ measured in 4 devices extensively studied in this work are collected in Fig. 3 and plotted on a logarithmic scale for both electrons and holes (no electron-hole asymmetry was noticed). The plot covers the experimental range of $|n|$ from $10^9$ to nearly $10^{12}$ cm$^{-2}$. All the data fall within the range marked by the two dashed curves that correspond to constant $v_F = v^*_F$ and $v_F = 3\times10^6$ m/s. One can see a gradual increase in $v_F$ as $n$ increases, although the logarithmic scale makes the observed threefold increase less dramatic than in the linear presentation of Fig. 2c.

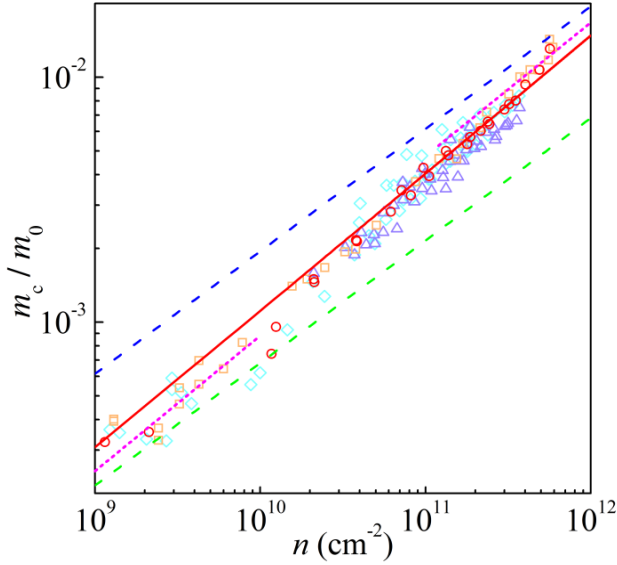

FIG. 3. Interaction-induced changes in the cyclotron mass. Different symbols are the measurements for different devices. Blue and green dashed lines are the behavior expected for the linear spectrum with constant $v_F$ equal to $v^*_F$ and $3\times10^6$ m/s, respectively. $m_0$ is the free electron mass. The solid red curve is for the spectrum renormalized by e-e interactions and described by eq. (2) that takes into account the intrinsic screening self-consistently. The two dotted curves show that the interaction effects can also be described by a simpler theory (eq. 1) with an extra fitting parameter $\varepsilon_G(n)$, graphene's intrinsic dielectric constant. The best-fit curves yield $\varepsilon_G \approx 2.2$ and 4.9 at low and high ends of the $n$ range, in reasonable agreement with the values expected in the RPA.

Even for the highest $n$ in Fig. 3, the measured $m_c$ do not reach the values expected for $v_F = v^*_F$ and are better described by $v_F \approx 1.3\, v^*_F$. This can be due to the fact that the highest $n$ we could achieve for suspended graphene were still within a sub-$10^{12}$ cm$^{-2}$ range, in which some enhancement in $v_F$ was reported for graphene on SiO$_2$ [14]. Alternatively, the difference could be due to the absence of a substrate in our case. To find out which of the effects dominates, we have studied high-μ devices made from graphene deposited on boron nitride [20,21] (its dielectric constant ε is close to that of SiO$_2$) and found that $m_c$ in the range of $n$ between ≈0.1 and $1\times10^{12}$ cm$^{-2}$ is well described by $v_F \approx v^*_F$ [18]. This indicates that the observed difference in $m_c$ at high $n$ in Fig. 3 with respect to the values expected for $v^*_F$ is likely to be due to the absence of dielectric screening in suspended graphene, which maximizes the interaction effects.

To explain the observed strong changes in $v_F$, let us first note that, in principle, not only e-e interactions but other mechanisms such as electron-phonon coupling and disorder can also lead to changes in $v_F$. However, the fact that the increase in $v_F$ is observed over a wide range of low $E$ rules out electron-phonon mechanisms whereas the virtual absence of disorder in our suspended graphene makes the influence of impurities also unlikely. Therefore, we focus on e-e interactions, in which case graphene's spectrum is modified as shown in Fig. 1 and, in the first approximation, can be described by two related equations [8-10].

$$\frac{k}{v_F}\frac{\partial v_F}{\partial k} = -\frac{e^2}{4\varepsilon\hbar v_F} \qquad (1)$$

$$\frac{k}{v_F}\frac{\partial v_F}{\partial k} = -\frac{2}{\pi^2}\left[1 - \frac{4\hbar v_F \varepsilon}{N_f e^2} + \frac{8\hbar v_F \arccos\left(\frac{\pi N_f e^2}{8\hbar v_F \varepsilon}\right)}{N_f e^2 \pi \sqrt{1-\left(\frac{\pi N_f e^2}{8\hbar v_F \varepsilon}\right)^2}}\right] \qquad (2)$$

where $\varepsilon =(1+\varepsilon_s)/2$ describes the effect of a substrate with a dielectric constant $\varepsilon_s$. Equation (1) can be considered as the leading term in the RGT expansion in powers of $\alpha = e^2/\varepsilon\hbar v_F$ whereas (2) corresponds to a similar expansion in powers of $1/N_f$ [8-10]. The diagrams that depict these approximations are given in [18]. Importantly, eq. (2) self-consistently includes the screening by graphene's charge carriers. An approximate scheme to incorporate this intrinsic screening while keeping the simplicity of eq. (1) is to define an effective screening constant $\varepsilon_G(n)$ for the graphene layer and add it to ε (for suspended graphene ε = $\varepsilon_G$). Then, integrating eq. (1), we obtain the logarithmic dependence [8]

$$v_F(n) = v_F(n_0)\left[1 + \frac{\alpha}{8\varepsilon_G}\ln(n_0/n)\right] \qquad (3)$$

where $n_0$ is the concentration that corresponds to the ultraviolet cut-off energy Λ, and $v_F(n_0)$ is the Fermi velocity near the cut-off. We assume $v_F(n_0) \equiv v^0_F$, its accepted value in graphene structures with week e-e interaction.

Both approximations result in a similar behavior of $v_F(n)$ and provide good agreement with the experiment. However, eq. (2) is more general, self-consistent and essentially requires no fitting parameters because Λ is expected to be of the order of graphene's bandwidth and affects the fit only weakly, as log(Λ). Alternatively, Λ can be estimated from the known value of $v^0_F$ at high $n \approx 5\times10^{12}$ cm$^{-2}$ as Λ =2.5±1.5 eV [22]. The solid curves in Figs. 2b,c and 3 show $m_c(n)$ and $v_F(n)$ calculated by integrating eq. (2) and using Λ ≈3eV. The dependence captures all the main features of the experimental data. As for equations 1 and 3, they allow a reasonable fit by using $\varepsilon_G$ ~3.5 over the whole range of our $n$. More detailed analysis (dotted curves in Fig. 3) yields $\varepsilon_G \approx 2.2$ and 5 for $n$ ~$10^9$ and $10^{12}$ cm$^{-2}$, respectively. These values are close to those calculated in the random phase approximation (RPA) which predicts $\varepsilon_G = 1 + \pi N_f e^2/8\hbar v_F$. By using this expression in combination with eq. (3) leads to a fit that is practically indistinguishable from the solid curve given by eq. (2). This could be expected because eq. (2) includes the screening self-consistently, also within the RPA. The value of $\varepsilon_G$ has recently become a subject of considerable debate [23-27]. Our data clearly show no anomalous screening, contrary to the recent report [26] that suggested $\varepsilon_G \approx 15$.

Finally, a large number of theories have been predicting that the diverging contribution of e-e interactions at low $E$ may result in new electronic phases [28-30], especially in the least-screened case of suspended graphene with ε =1. Our experiments shows the diverging behavior of $v_F$ but no new phases emerge, at least for $n >10^9$ cm$^{-2}$ ($E >$4meV). Moreover, we can also conclude that there are no insulating phases even at $E$ as low as 0.1 meV. To this end, we refer to Ref. [18] in which we present the data for graphene's resistivity $\rho(n)$ in zero $B$. The peak at the NP continues to grow monotonically down to 2 K, and $\rho(T)$ exhibits no sign of diverging (the regime of smearing by spatial inhomogeneity is not reached even at this $T$). This shows that, in neutral graphene in zero $B$, there is no gap larger than ≈0.1 meV. This observation is consistent with the fact that $v_F$ increases near the NP, which leads to smaller and smaller $\alpha = e^2/\hbar v_F$ at low $E$ and, consequently, prevents the emergence of the predicted many-body gapped states.

In conclusion, "the linear Dirac-like spectrum of graphene" is no longer linear if one refers to a wide range of $n$, especially below $10^{11}$cm$^{-2}$. The spectral slope (that is, the Fermi velocity) can change by a significant factor, which can be explained by e-e interactions. The RGT provides a good description for the experimentally observed spectrum, even though the measured changes are not small perturbations. The experimental observations reported here show that graphene is a unique example of "marginal Fermi liquid behavior" [31] in condensed matter physics.

REFERENCES

[1] R. Shankar, *Rev. Mod. Phys.* **66**, 129 (1994).
[2] V. N Kotov *et al.*, arXiv:1012.3484 (2010).
[3] D. V. Khveshchenko, *Phys. Rev. Lett.* **87**, 246802 (2001).
[4] E. V. Gorbar *et al.*, *Phys. Rev. B* **66**, 045108 (2002).
[5] J. E. Drut and T. A. Lahde, *Phys. Rev. Lett.* **102**, 026802 (2009).
[6] W. J. Liang *et al.*, *Nature* **411**, 665 (2001).
[7] A. A. Abrikosov and D. Beneslavskii, *Sov. Phys. JETP* **32**, 699 (1971).
[8] J. Gonzalez, F. Guinea and M. A. H. Vozmediano, *Nucl. Phys. B* **424**, 595 (1994).
[9] J. Gonzalez, F. Guinea and M. A. H. Vozmediano, *Phys. Rev. B* **59**, 2474 (1999).
[10] M. S. Foster and I. L. Aleiner, *Phys. Rev. B* **77**, 195413 (2008).
[11] K. S. Novoselov *et al.*, *Nature* **438**, 197 (2005).
[12] Y. B. Zhang *et al.*, *Nature* **438**, 201 (2005).
[13] C. L. Kane and E. J. Mele, *Phys. Rev. Lett.* **93**, 197402 (2004).
[14] Z. Q. Li *et al.*, *Nature Phys.* **4**, 532 (2008).
[15] X. Du *et al.*, *Nature Nanotech.* **3**, 491 (2008).
[16] K. I. Bolotin *et al.*, *Solid State Commun.* **146**, 351 (2008).
[17] E. V. Castro *et al.*, *Phys. Rev. Lett.* **105**, 266601 (2010).
[18] See supplementary material at http:// .
[19] S. G. Sharapov, V. P. Gusynin and H. Beck, *Phys. Rev. B* **69**, 075104 (2004).
[20] C. R. Dean, *et al.*, *Nature Nanotech.* **5**, 722 (2010).
[21] D. A. Abanin *et al.*, *Science*, in press (2011).
[22] F. de Juan, A. G. Grushin, and M. A. H. Vozmediano, *Phys. Rev. B* **82**, 125409 (2010).
[23] Y. Barlas *et al.*, *Phys. Rev. Lett.* **98**, 236601 (2007).
[24] E. H. Hwang, B. Y. K. Hu and S. Sarma, *Phys. Rev. Lett.* **99**, 226801 (2007).
[25] D. E. Sheehy and J. Schmalian, *Phys. Rev. Lett.* **99**, 226803 (2007).
[26] V. N. Kotov, B. Uchoa and A. H. C. Neto, *Phys. Rev. B* **78**, 035119 (2008).
[27] J. P. Reed *et al.*, *Science* **330**, 805 (2010).
[28] D. V. Khveshchenko, *Phys. Rev. Lett.* **87**, 246802 (2001).
[29] E. V. Gorbar *et al.*, *Phys. Rev. B* **66**, 045108 (2002).
[30] J. E. Drut and T. A. Lähde, *Phys. Rev. Lett.* **102**, 026802 (2009).
[31] C. M. Varma *et al.*, *Phys. Rev. Lett.* **63**, 1996 (1989

# SUPPLEMENTARY INFORMATION:

# Dirac cones reshaped by interaction effects in suspended graphene

D. C. Elias *et al*

#1. Experimental devices

Graphene monolayers were obtained by micromechanical cleavage of graphite on top of an oxidized Si wafer [S1]. In this work, we specially selected long and narrow crystals (typically, 2 to 4 µm wide) which allowed us to avoid dry etching of graphene mesas. Two-terminal devices such as shown in Fig. 2 of the main text were then designed and fabricated by using standard lithography and deposition techniques. The 300 nm $SiO_2$ layer was partially etched in a buffered HF solution to leave graphene hanging above the substrate. The metal leads (5 nm Cr followed by 100 nm of Au) remained not fully etched underneath and served as a mechanical support. These fabrication procedures are similar to those described in refs. S2-S5.

The current annealing was performed in situ, in a liquid-helium bath by applying voltage between adjacent contacts. Current densities of ~1 mA/µm were necessary to heat suspended graphene locally to $T > 600^{o}C$ [S5]. Our devices either fail or anneal after a minor (<1%) increase in applied voltage, which we believe is an indication that the real $T$ of annealing could be even higher than suggested in ref. [S5].

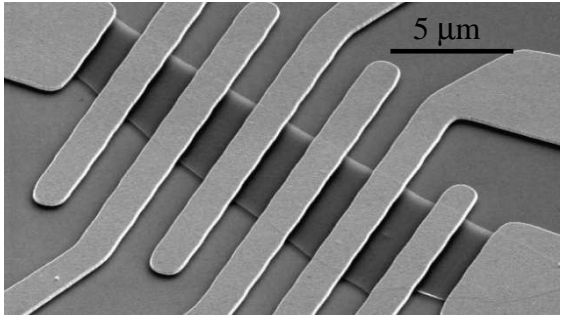
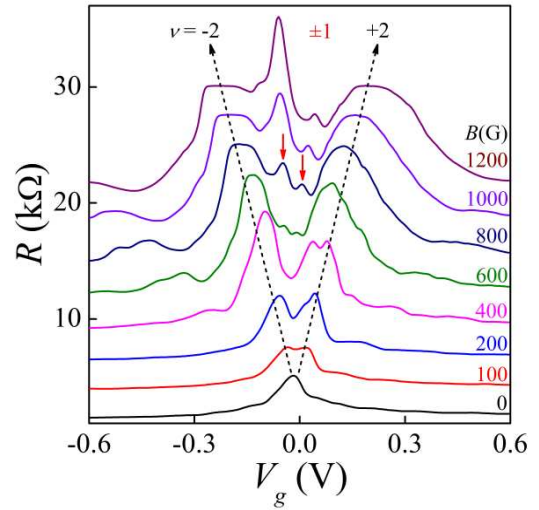

FIG S1. Our graphene devices. Left – Scanning electron micrograph of another suspended device, different from the one shown in Fig. 2a. Right – Typical behaviour of $R(V_g)$ measured at 2K. The curves are shifted for clarity. The QHE in the two probe geometry is known to lead to plateaux in $R$ at $h/\nu e^2$. Such QHE plateaux are clearly seen in our devices below 0.1T. The dominant QHE plateau (filling factor $\nu = \pm 2$) at $R \approx 12.8$kΩ is first formed at negative gate voltages where µ is somewhat higher. Additional peaks at lower $|V_g|$ correspond to $\nu = \pm 1$ and indicate either spin or valley splitting.

Figure S1 shows two-terminal resistance $R$ as a function of gate voltage $V_g$ in different magnetic fields $B$. We refer to our measurements as two-terminal because the supporting metal contacts overlap with the current path (Fig. S1), that is, they are invasive [S6,S7]. In this measurement geometry, we found little difference whether we used two- or four-probe measurement geometry because of the relatively small resistance of the metal leads.

As one can see in Figure S1, the Landau level splitting occurs at $B \sim 100$ G (red and blue curves). The observation of SdHO requires $\mu B \approx 1$, which allows us to estimate quantum mobility $\mu$ as $\sim 10^6$ cm$^2$/Vs [S3,S4,S8]. This value is in good agreement with the field-effect $\mu$ found from changes in conductivity $\sigma$ as a function of $n$ in zero $B$ [S4] (also, see Fig. S2). As a further indication of the graphene quality, one can see that the first quantum Hall effect (QHE) plateau develops at 600 G for holes (green curve; negative $V_g$) and becomes fully formed for both electrons and holes at 1000 G (violet). Also, the 4-fold degeneracy of the lowest LL becomes lifted already at $\sim 600$ G (green).

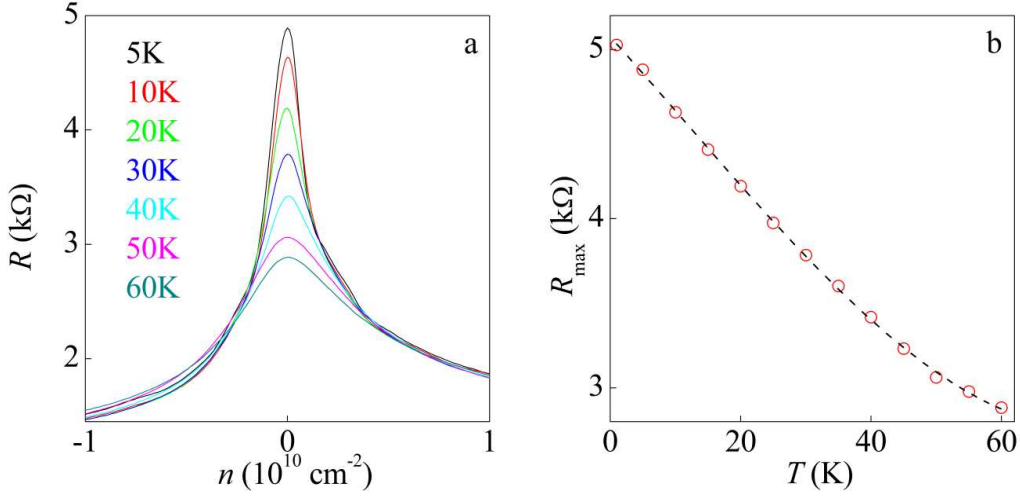

FIG S2. No discernable gap in neutral graphene. (**a**) – $R$ as a function of concentration $n$ in a suspended device at various $T$ in zero $B$. The peak at the Dirac point continues to sharpen with decreasing $T$ but $R$ remains finite, with no sign of a gap: that is, $R(T)$ does not diverge at $T \to 0$. (**b**) – The device's maximum resistance as function of $T$. The points are the experimental data and the dashed curve is a guide to the eye. The practically linear dependence $R(T)$ is puzzling and may be related to the transition from the dependence $R \propto 1/T^2$ found at high $T$ (due to thermally generated carriers at the NP) to the pseudo-diffusive regime with a finite conductivity in the limit of low $T$.

Charge inhomogeneity $\delta n$ is usually estimated from smearing of the resistance peak near the NP. However, in our devices, the peak continues sharpening down to 2 K (Fig. S2), the lowest $T$ in the current experiments. This shows that the thermal generation of electrons and holes at the NP dominates any remnant charge inhomogeneity, which yields $\delta n$ less than $\sim 10^8$ cm$^{-2}$, that is of about one electron per square $\mu$m. In order to extract cyclotron mass $m_c$ it was necessary to measure SdHO at many different $T$. This effectively led to $\delta n$ being determined by $T$ rather than real inhomogeneity and limited our $m_c$ measurements to $n \geq 10^9$ cm$^{-2}$. Furthermore, the smooth monotonic behaviour of $R$ as a function of both $n$ and $T$ (see Fig. S2) implies that, except for the discussed logarithmic corrections, no dramatic reconstruction of the Dirac spectrum occurs at $E$ down to 1 meV ($n \approx 10^8$ cm$^{-2}$). Otherwise, one would expect to observe some anomalies in $R(n,T)$ whereas the presence of an energy gap larger than $\sim 0.1$ meV would be seen as diverging $R(T \to 0)$.

#2. Analysis of Shubnikov–de Haas oscillations

We have measured the cyclotron mass $m_c$ in graphene by analysing $T$ dependence of SdHO. This well-established approach is widely used in literature and, in the case of graphene, provided accurate measurements of $m_c$ which have later been found in good agreement with the results obtained by other techniques (e.g., magneto-optics and tunnelling microscopy). In brief, our procedures involved measurements of suspended graphene's conductance $G = 1/R$ as a function of $n$ at a given $B$. Then, we changed $T$ and repeated the measurements. $T$ and $B$ were always chosen to keep far away from the QHE regime so that changes in conductance $\Delta G \ll G$.

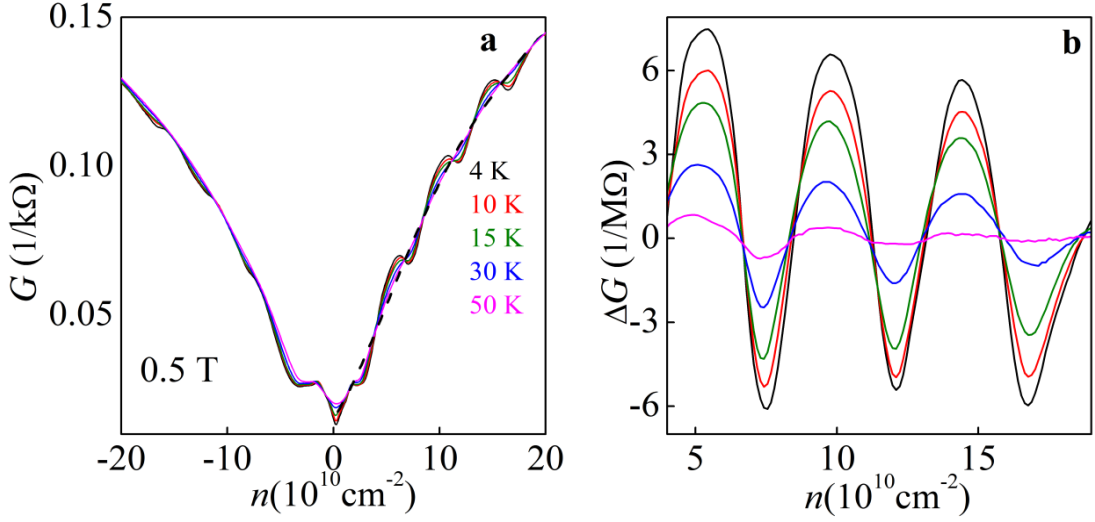

FIG S3. (a) – $G(n)$ for a suspended graphene device in $B = 0.5$ T at several $T$. The dashed curve indicates the smooth polynomial background. (b) – Curves from (a) after the subtracting the background.

Examples of our raw data are shown Figure S3a. SdHO are clearly seen on top of the standard V-shaped background. This background is smooth and, for easier analysis, can be subtracted. We have done this separately for electrons and holes. To standardise the procedures, we normally defined the background by fitting a 4th-order polynomial to one of high-$T$ curves $G(n)$ with no discernable oscillations, as illustrated in Fig. S3a. The subtraction resulted in curves such as shown in Fig. S3b. The SdHO amplitude was then calculated as the difference between $\Delta G$ in maxima and minima. This yielded the data such as shown in Fig. 2a of the main text. Typically, we used 10 different $T$ to obtain each value of $m_c$. The results were practically independent of the choice of subtracted background and other procedural details, essentially because we analyzed the difference between minima and maxima.

#3. Influence of a dielectric substrate

As found in many experiments, graphene on $SiO_2$ exhibits the Fermi velocity $v^*_F \approx 1.05 \pm 0.1 \times 10^6$ m/s for the typically accessible range of $n \sim 10^{12}$ cm$^{-2}$. The measurements reported in the main text show a slightly higher $v_F$ (15 to 25%) in suspended graphene for the same range of $n$. This disagreement can be attributed to the absence of dielectric screening in the suspended devices. To prove this and exclude any systematic error arising due to the use of devices with drastically different mobilities ($\mu$ differ by a factor of 100 for suspended graphene and graphene on $SiO_2$), we performed measurements of $m_c(n)$ for graphene on boron nitride (GBN). The latter devices allow $\mu > 100,000$ cm$^2$/V and, at the same time, e-e interactions are screened in a manner similar to the case of graphene on $SiO_2$ (boron nitrite exhibits $\varepsilon_s \approx 5$ [S9]).

Our GBN devices were fabricated as described in refs. [S10,S11] and one of the studied devices is shown in Fig. S4. To find $m_c$, we performed the same measurements and analysis as described in the previous chapter. The resulting dependence $m_c(n)$ is shown in Fig. S4. The accessible range of $n$ was limited to $\geq 10^{11}$ cm$^{-2}$ due to charge inhomogeneity that was smaller than in graphene on $SiO_2$ but still significant, in agreement with the results of refs. [S11,S12]. The dashed curve corresponds to a constant $v_F = v^*_F$ and provides an excellent description of our data within this limited range of *n*, similar to the case of graphene on $SiO_2$. This strongly supports the argument that $v_F$ in graphene on a substrate is lower than in suspended graphene due to dielectric screening in the former case.

To check our analysis of the renormalized spectrum for consistency, the solid and dotted curves in Figure S4 show $m_c(n)$ calculated by using to equation (2). The dotted line is the same theory curve shown in Figs. 2c and 3 of the main text for suspended graphene, which corresponds to the case of $\varepsilon = 1$ and $\Lambda \approx 3$eV. On the other hand, the solid line was calculated by using the same equation and only adding the dielectric screening due to boron nitride with no change in other parameters. The agreement between the experiment and theory is impressive and shows that our theoretical description is able to explain not only the *n* dependence of the Fermi velocity but, also, its dependence on dielectric screening.

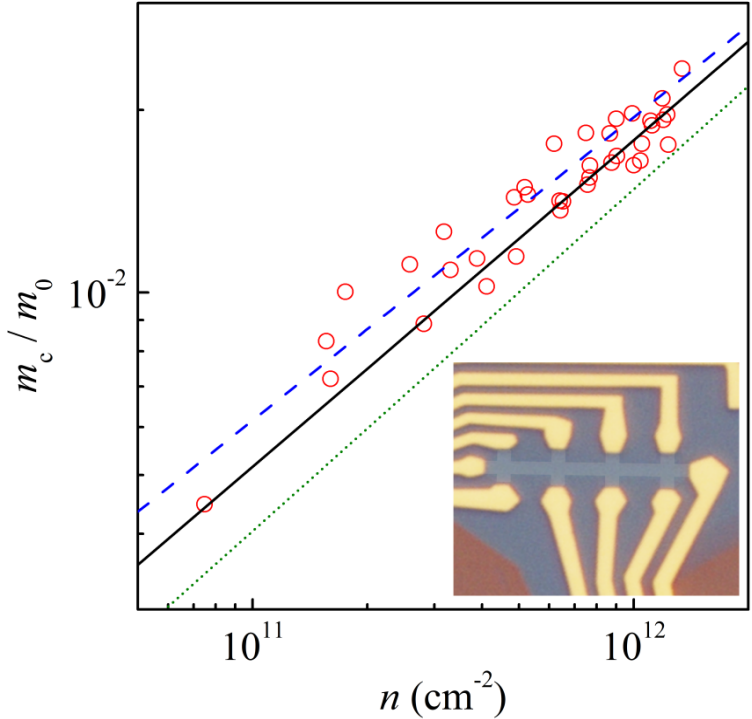

FIG S4. Cyclotron mass as function of *n* for graphene on boron nitride. The symbols are experimental data; the dashed line is the non-interacting behaviour with constant $v_F = v^*_F$. The RGT approach, which is used in the main text to describe $m_c(n)$ in suspended graphene over a wide range of *n*, is also consistent with the limited-range data for GBN devices. The dotted curve is given by equation (2) of the main text ($\varepsilon=1$; $\Lambda=3$eV) whereas the solid one is for $\varepsilon_s=5$; $\Lambda=3$eV (no fitting parameters). The inset shows an optical micrograph of a Hall bar device made from graphene deposited on BN (no encapsulating top layer [S11]). For clarity, the contrast of the 1μm wide graphene mesa was digitally enhanced.

#### #4. Interaction renormalization of the Dirac spectrum in various approximations

Near the NP, screening is weak due to the low density of states and completely suppressed in neutral graphene because the density of states goes to zero. As a result, electronic levels become increasing affected by e-e interactions as their energy approaches the Dirac point. The Hartree-Fock correction to the quasiparticle energy is given by

$$\delta E \approx \pm \int d^2\vec{k}' \frac{2\pi e^2}{\varepsilon |\vec{k}-\vec{k}'|} \left( \frac{1}{2} \pm \frac{\vec{k}\cdot\vec{k}'}{|\vec{k}||\vec{k}'|} \right) \approx \pm \frac{e^2}{4\varepsilon} \ln\left(\frac{k_\Lambda}{|\vec{k}|}\right) \quad (S1)$$

where $k_\Lambda$ is the upper limit in the momentum integral, and the signs ± correspond to electrons and holes, respectively. This equation yields a change in the Fermi velocity $\delta v_F$ which becomes a function of momentum $\vec{k}$

$$\delta v_F \approx \frac{e^2}{4\varepsilon} \ln\left(\frac{k_\Lambda}{|\vec{k}|}\right) \quad (S2).$$

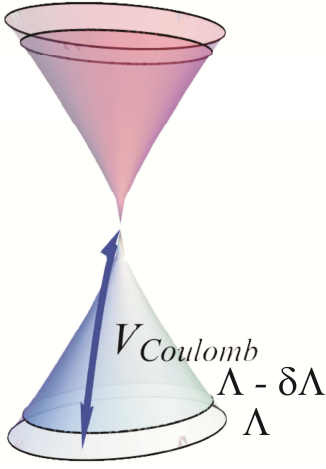

FIG S5. Sketch for the Renormalization Group procedure used to explain the experimental observations. Coulomb interactions between low- and high- $E$ states deplete the electronic spectrum near the Dirac point.

An improvement over the Hartree-Fock approximation can be achieved by calculating changes in $v_F$ for low-$E$ quasiparticles, which are induced by their interaction with high-$E$ excitations in the interval of energies $\Lambda - \delta\Lambda \leq E \leq \Lambda$ and defining a new model for the electronic spectrum in which these excitations are removed, as schematically shown in Fig. S5. Within this model, $v_F$ is described by

$$\hbar v_F(\Lambda - \delta\Lambda) \approx \hbar v_F(\Lambda) + \frac{e^2}{4\varepsilon} \frac{\delta\Lambda}{\Lambda} \quad (S3)$$

Or, alternatively

$$\hbar k \frac{\partial v_F}{\partial k} = -\frac{e^2}{4\varepsilon} \quad (S4).$$

This result reproduces equation (1) in the main text. Using the same analysis, it can be shown that there is no need to modify other parameters in the Hamiltonian. This scheme defines the RGT transformation that is exact in the limit $\alpha = e^2/\hbar v_F << 1$. The self energy diagram that gives rise to eq. (1) is shown in Fig. S6a.

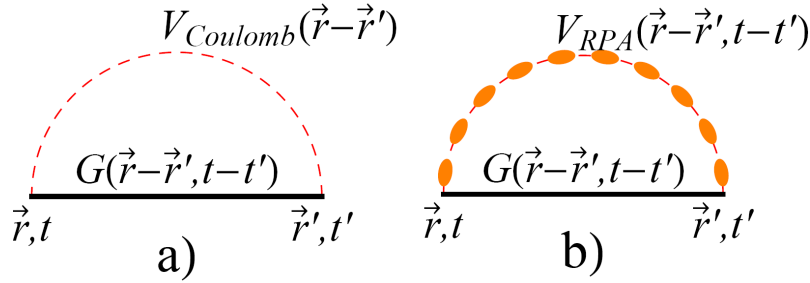

FIG S6. (a) – Diagram that leads to eq. (1) of the main text. (b) –The diagram takes into account self-screening.

Equations (1) and (S1-S4) include only screening effects due to environment of the graphene sheet, which is described by the dielectric constant $\varepsilon$. The intrinsic screening by charge carriers can also be added in a phenomenological way by redefining $\varepsilon$ and introducing $\varepsilon_G$ as discussed in the main text. Alternatively, a better description can be achieved by self-consistently including the screening processes into the interaction line in Fig. S6a. The resulting diagram is shown in Fig. S6b, and this leads to equation (2) of the main text. Furthermore, it can be shown that the infinite summation of polarization bubbles in the second diagram results in the approximation that becomes exact if $N_f >> 1$. In graphene, $N_f = 4$ so that the approximation's accuracy is comparable to similar calculations used in quantum chromodynamics [S13].

#5. Influence of disorder

The RGT flow that describes the dependence of $v_F$ on energy leads to changes in this parameter, which can be comparable to $v^0_F$, the initial values of the parameter itself. On the other hand, other couplings such as electron-phonon [S14] and electron-plasmon interactions [S15] can be treated within a perturbation theory because they do not lead to logarithmic divergences. Therefore, it can be expected that their effect on the Fermi velocity does not exceed a fraction of its value and, accordingly, they cannot explain the large enhancement observed in the experiment. The only other interaction that can lead to logarithmic renormalization is the coupling to some types of scalar and gauge random disorder [S16-S18]. However, the arising corrections have the opposite sign with respect to that due to electron-electron interactions. Furthermore, the disorder can be described by the dimensionless parameter

$$\Delta \sim \langle V^2 \rangle (l/v_F)^2 \qquad (S5)$$

where $\langle V^2 \rangle$ gives the average value of the disorder, and $l$ is the range over which it is correlated. This gives rise to a scattering time $\tau$

$$h/\tau \sim \Delta \times E_F \qquad (S6)$$

where $E_F$ is the Fermi energy. In order to significantly change the effect of electron-electron interaction, the value of $\Delta$ should be comparable to $e^2/\hbar v_F$. The long mean free path, characteristic of the suspended graphene studied in this work, rule out the existence of such strong disorder.

Supplementary References

[S1]  K. S. Novoselov *et al*. *PNAS* **102**, 10451 (2005).
[S2]  X. Du *et al*. *Nature Nanotech.* **3** 491 (2008).
[S3]  K. I. Bolotin *et al*. *Solid State Commun.* **146** 351 (2008).
[S4]  E. V. Castro et al. *Phys. Rev. Lett.* **105**, 266601 (2010).
[S5]  K. I. Bolotin et al. *Nature* **462** 351 (2008).
[S6]  P. Blake *et al*. *Solid State Commun* **149**, 1068 (2009)
[S7]  J. R. Williams *et al*. *Phys. Rev. B.* **80**, 045408 (2009)
[S8]  M. Monteverde *et al*. *Phys. Rev. Lett.* **104**, 126801 (2010).
[S9]  R. V. Gorbachev *et al*. *Small* **7**, 465 (2011).
[S10] C. R. Dean *et al*. *Nature Nanotech.* **5**, 722 (2010).
[S11] A. S. Mayorov *et al*. arXiv:1103.4510
[S12] J. Xue *et al*. *Nature Mater.* **10**, 282 (2011)
[S13] S. Coleman, *Aspects of Symmetry*, Cambridge University Press (1985).
[S14] C.-H. Park et al. *Phys. Rev. Lett.* **99**, 086804 (2007).
[S15] A. Bostwick et al. *Science* **328**, 999 (2010).
[S16] T. Stauber et al. *Phys. Rev. B* 71, 041406, (2005).
[S17] M. S. Foster and I. L. Aleiner, *Phys. Rev. B*, **77**, 195413 (2008).
[S18] I. F. Herbut et al. *Phys. Rev. Lett.* **100**, 046403 (2008).